\documentclass{article}
\usepackage[utf8]{inputenc}
\usepackage{graphicx}
\usepackage{amsmath}
\usepackage[danish, english]{babel}
\usepackage{subfig}
\usepackage{breqn}
\usepackage{fancyhdr}
\usepackage{lastpage}
\usepackage{float}

\usepackage[numbers]{natbib}
\usepackage{url}
\newcommand{\appropto}{\mathrel{\vcenter{
  \offinterlineskip\halign{\hfil$##$\cr
    \propto\cr\noalign{\kern2pt}\sim\cr\noalign{\kern-2pt}}}}}

\begin{document}

\section*{Abstract}

Fast and accurate protein structure prediction is one of the major
challenges in structural biology, biotechnology and molecular biomedicine.
These fields require 3D protein structures for rational design of
proteins with improved or novel properties.
X-ray crystallography is the most common approach even with its low success rate, but lately NMR based approaches have gained popularity.
The general approach involves a set of distance restraints used to guide a structure prediction, but simple NMR triple-resonance experiments often provide enough structural information to predict
the structure of small proteins. Previous protein folding
simulations that have utilised experimental data have weighted the
experimental data and physical force field terms more or less
arbitrarily, and the method is thus not generally applicable to new
proteins. Furthermore a complete and near error-free assignment of chemical
shifts obtained by the NMR experiments is needed, due to the static, or deterministic, assignment.\\
In this thesis I present {\tt Chemshift}, a module for handling chemical shift assignments, implemented in the protein structure determination program {\tt Phaistos}. 
This module treats both the assignment
of experimental data, as well as the weighing compared to physical
terms, in a probabilistic framework where no data is discarded. Provided a partial
assignment of NMR peaks, the module is able to improve the assignment
with the intension to utilise this in the protein folding with little bias.

\section*{Acknowledgements}
I'd like to thank my supervisor Jan H. Jensen for not bullying me as much as he does other
students. Thanks to Casper S. Svendsen for inspiring me for future instructor work. Thanks
to Anders S. Christensen for being perfect in every way. Thanks to Qian for not killing us, and lastly thanks to Jimmy for his good sense of humour.

\newpage
\tableofcontents
\newpage

\section{Introduction}
\label{section:Introduction}

To generalise there have been three branches in protein structure determination.
X-ray crystallography is the most common approach, that gives very accurate structures and protein size is in general not an issue. It however has a very low success rate, since most proteins of interest does not easily crystallise.
Another less popular experimental approach involves using NMR data to create a set of Nuclear Overhauser Effect (NOE) distance restraints. From these restraints the protein structure can be deduced, but protein size is a limiting factor and structures can in general not be inferred from large proteins.
In the opposite end of the spectrum is the purely computational methods, that uses force fields to simulate protein mechanics. These methods uses a lot of approximations in order to provide results on a reasonable time-scale for large systems as proteins, which often hinder the correct conformers to be predicted.
The quality of predictions from computational methods have recently been improved by including experimental chemical shifts alongside with force fields and chemical shift predictors in the structure prediction \citep{MeilerBaker,Vendruscolo,Bax}. \\
A necessary step between experiments and determining the protein structure is assignment of the measured chemical shifts, which for larger proteins can be very time consuming and is a major bottleneck.
Several methods have been developed to automate this \citep{Garant,MARS,TATAPRO}, but most still require a great deal of human intervention. Two methods that require minimal intervention is Autoassign \citep{Autoassign} and FLYA \citep{FLYA}.
The strengths of Autoassign is that it is a free service and that chemical shifts are analysed and assigned very quickly (typically less than a minute) with few wrong assignments. FLYA has been shown to perform better than Autoassign, but is slower and requires a license to use.\\
A 2003 study estimated that 40\% of all proteins in the Biological Magnetic Resonance Data Bank \citep{BMRB} (BMRB) contain at least one mis-assigned chemical shift \citep{RefDB}. The more severe errors might affect the predicted structures, since data is discarded if the structure calculations don't converge. And even non-erroneous assignments might restrict the predicted conformers in cases where a protein has more than one native conformation.\\

The purpose of this work is as follows:
\begin{itemize}
    \item Remove the need for a manual assignment.
    \item Derive an energy function based on Bayesian inference principles for describing experimental data.
    \item Implement in the protein structure prediction program {\tt Phaistos} a probabilistic method to include experimental data in structure prediction.
    \item Allow the assignment of chemical shifts to change during structure prediction, without discarding data.
    \end{itemize}

In this thesis the current state of the development of the {\tt Chemshift} module in {\tt Phaistos} is presented. Emphasis has been put on keeping the thesis short and readable, while presenting details of background, theory and computational implementation to an extent such that the thesis, along side with the code itself, can be used to maintain or recreate the module.\\

To avoid any confusion, throughout the thesis a peak will refer to the chemical shifts from two or three linked nuclei. A spin system is the linked nuclei which give rise to a peak in the NMR spectrum. A spin system array is computationally the array that holds the assignment of peak. Each array belong to a specific type of experiment and spin system. When differences of chemical shifts is mentioned, only differences between chemical shifts from the same nuclei is assumed.

\section{Background}
\label{section:Background}
In atomic nuclei isotopes with non-zero magnetic moments, an energy difference due to Zeeman-splitting is observed between the different spin-states when a strong external magnetic field is applied. The local magnetic field these nuclei experience is slightly perturbed (shielded) by the local molecular environment, which causes the local environment to be reflected in the size of the energy-splitting. \\

With Nuclear Magnetic Resonance (NMR) spectroscopy, the resonance frequency $\nu$ of the nucleus can be measured. But since this frequency is dependent of the field used, it is convenient to relate this to a reference frequency $\nu_{ref}$ as \citep{Sauer}

\begin{equation}
\delta = 10^6 \frac{\nu-\nu_{ref}}{\nu_{ref}},
\end{equation}
where $\delta$, in units of ppm, is called the chemical shift.

\begin{figure}
    \centering
      \includegraphics[width=0.6\textwidth]{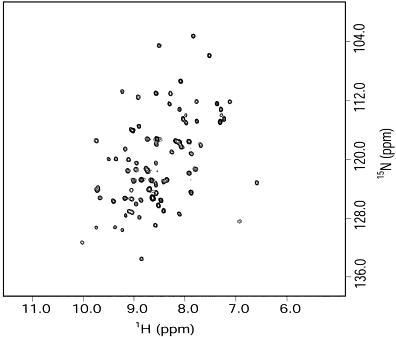}
      \caption{Contour plot of the $^1H$–$^{15}N$ HSQC spectrum of recombinant human ubiquitin encapsulated in AOT reverse micelles dissolved in n-pentane \citep{Venters}}
      \label{fig:hsqc}
\end{figure}

By utilising the coupling between neighbouring nuclei in a protein, one can correlate a nuclei chemical shift with another. One example is the two-dimensional HSQC-experiment which correlates a $^{15}$N nuclei with the neighbouring $^1$H nuclei and thus a peak for every H-N pair can be observed (See Figure~\ref{fig:hsqc} for an example).

\begin{figure}[htb]
  \centering
  \subfloat[HSQC]{%
    \includegraphics[width=.24\textwidth]{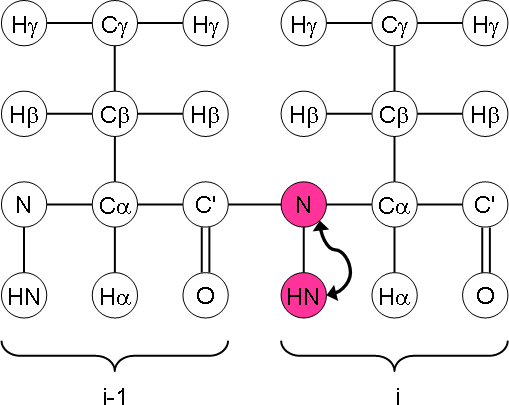}}\hspace{5pt}
  \subfloat[HNcaCO]{%
    \includegraphics[width=.24\textwidth]{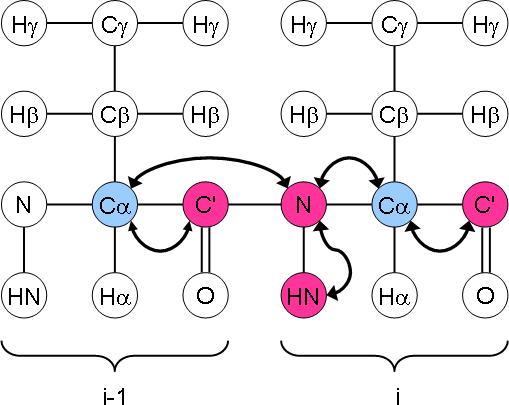}}\hspace{5pt}
  \subfloat[HNCA]{%
    \includegraphics[width=.24\textwidth]{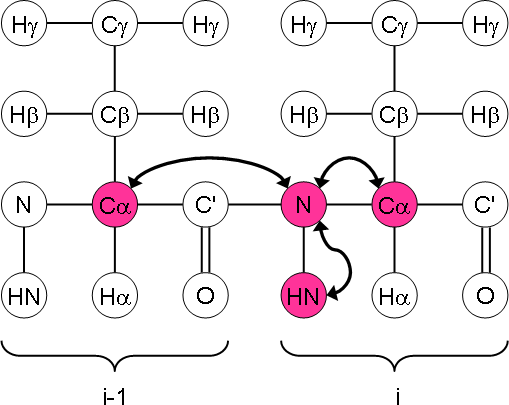}}\\
  \subfloat[HNcoCA]{%
    \includegraphics[width=.24\textwidth]{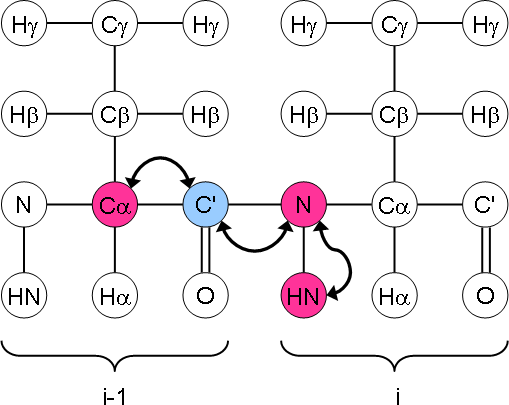}}\hfill
  \subfloat[HNcoCACB]{%
    \includegraphics[width=.24\textwidth]{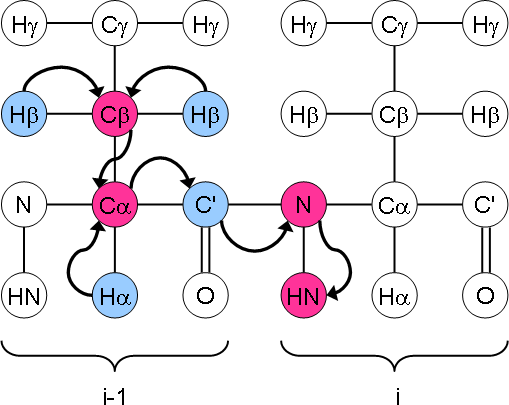}}\hfill
  \subfloat[HNCACB]{%
    \includegraphics[width=.24\textwidth]{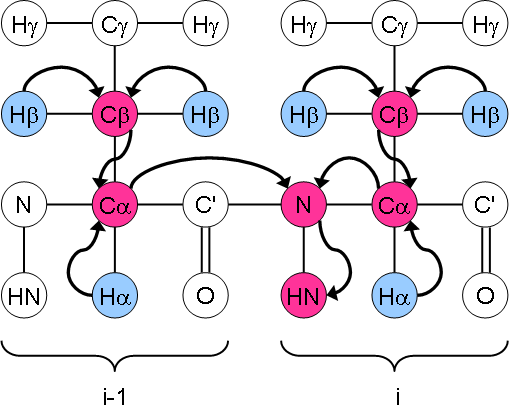}}\hfill
  \subfloat[HNCO]{%
    \includegraphics[width=.24\textwidth]{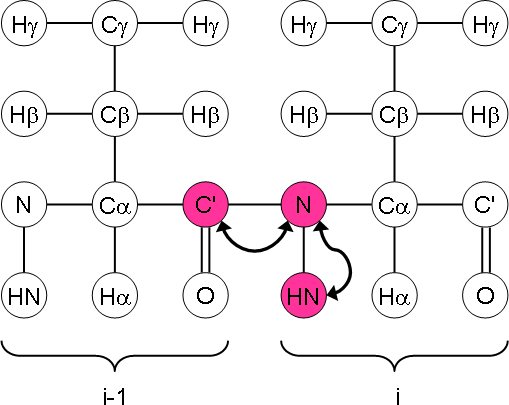}}\\
  \caption{The subfigures show which spin systems produces a resonance peak in each experiment \citep{Higman:proteinNMR}.}
  \label{fig:spinsystems}
\end{figure}

Several three-dimensional experiments can be performed as well. The most common ones couple H and N in a residue with one or more carbon nuclei from the same residue (refered to as intra or $i$), the preceding residue (inter or $i-1$) or both intra and inter.
Seven of the NMR experiments often used in backbone chemical shift assignment are shown in Figure~\ref{fig:spinsystems} for reference.

\clearpage
\newpage

\section{General Assignment Strategy}
\begin{figure}
    \centering
    \includegraphics[width=0.85\textwidth]{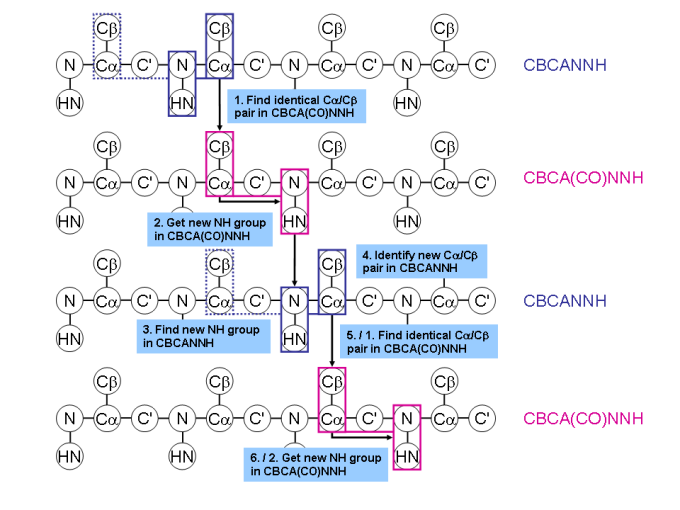}
    \caption{Depiction of how matching of chemical shifts can be used to establish a ladder of peaks which corresponding residues must precede each other in the protein. CBCANNH and CBCA(CO)NNH are synonyms for HNCACB and HNcoCACB respectively \citep{Higman:proteinNMR}}
    \label{fig:assign_theory}
\end{figure}

The NMR spectra contain no direct information about which residue each peak originates from. However using several experiments that probe different spin systems, it is possible to match identical chemical shifts in each experiment to the same nuclei. Furthermore inter and intra peaks can be matched together to form a ladder of chemical shift, as shown in Figure~\ref{fig:assign_theory}, only broken by Proline which doesn't have a H-N pair and therefore are not represented in these spectra.

This is of course not as easy as it sounds since there might be overlapping peaks in the spectra, strong redundancy at a specific chemical shift value, missing peaks or peaks originating from noise or impurities etc..
When the prementioned ladders are formed, it is often possible to assign these uniquely to a part of the protein. This is possible since especially CA and CB chemical shifts contain information about which amino-acid they originate from. Protein databases such as the Biological Magnetic Resonance Data Bank \citep{BMRB} (BMRB) can be used to collect statistics about chemical shifts from each amino-acid which can be used to infer the likelihood of the assignment. (See Appendix for an example)

When the spectra become more complex, for example with increased protein size, the assignment of the chemical shifts becomes increasingly more difficult, and in general complete assignments can't be constructed and erroneous assignments might be made.
A probabilistic framework can potentially remove the need for near 100\% certainty in an assignment. The general idea of probabilistic methods is that sparse data is better than no data, and as explained in the introduction, the ability to change the assignment of chemical shifts during protein folding are important for two major reasons. Errors from using a deterministic assignment have less impact, and you get more information from an incomplete assignment than you otherwise would.

\subsection{Select Automated Assignment Methods}
Two of the automated assignment methods that require the least human intervention is FLYA and Autoassign. This makes them suitable to use as alternatives to a manual assignment in the structure prediction, but they also provide a nice way to test how well an energy function describe these assignments of the data.

\subsubsection{Autoassign}
The general assignment strategy of Autoassign \citep{Autoassign} is to apply corrections to the chemical shift reference in each spectrum, to improve "between-spectra" alignment. Then peaks from the 3D spectra, with H and N chemical shifts within a set tolerance, is mapped to peaks in the HSQC-spectrum, to create pseudo-residues with all intra- and intermolecular nuclei mapped to a base N-H pair.\\
Peaks in HNCO with no corresponding peak in the HSQC spectrum, is used as a base and the previous step is repeated with these. It is argued that pseudo-residues which stems from side chain N-H pairs have low intensities in 3D experiments and thus pseudo-residues including less than three peaks from 3D spectra are recognised as side chains and are removed from backbone assignment. \\
If more pseudo-residues are created than there are assignable residues in the protein, the pseudo-residues with weakest intensities are set aside. And the $C^\alpha$ and $C^\beta$ peaks in these pseudo-residues are used to create amino-acid probability scores. \\
The most complete (containing most peaks) pseudo-residues intra and inter-peaks are paired and matched by a matching function. 
If the match is good and their combined amino-acid probability scores match a unique part of the sequence, the assignment is made. This is repeated with increasing tolerances until a full assignment is made or a upper bound on the constraints are reached. For the last step, the weaker pseudo-residues set aside earlier is analysed and assigned to one of the remaining missing residues if applicable or used to replace an already assigned one if it provide a better match.\\
The Autoassign article reports ~98\% of backbone chemical shifts being assigned for 7 proteins below ~150 residues in size with an error rate of 0.5\%, using 9 different NMR spectra.

\subsubsection{FLYA}
The assignment strategy in FLYA \citep{FLYA} is a mixture of deterministic and probabilistic approaches. A set of expected peak values is created based on sequence and chemical shift statistics. Each expected peak can be matched to only one experimental peak, but each experimental peak can be assigned multiple times. However if more peaks is found in a spectrum than 1.5 times the expected amount of peaks, the peaks with weakest intensities are removed. \\
A scoring function to evaluate the quality of the assignment is used together with an evolutionary algorithm to find the best assignment. No mathematical basis for the scoring function is given, but the gist of their approach is that an "external" part and an "internal" part contributes to the score with certain hand-picked weights. The external part evaluates how well the expected chemical shift value agrees with the mean value of the chemical shifts assigned to the nuclei. The internal part evaluates the variance of the assigned peaks. This evaluation is based on a normal distribution where a discrepancy of less than 1.5 and 2.0 times some predefined standard deviation for the external and internal part respectively, will contribute positively to the score, while discrepancies higher than this will favor that the assignment isn't made.\\
The FLYA article reports 96-99\% of backbone chemical shifts being assigned for three 100-150 residue proteins. A very large amount of NMR spectra was used, including NOE's, but instead of manually picking the peaks from these spectra, the peaks were automatically picked by other programs.

\section{Theory}
\label{section:Theory}
As mentioned previously, chemical shifts carry information about the protein structure, such as dihedral angles, side chain angles, ring current effects etc.. In the past chemical shifts have been used in a protein folding context, usually together with Nuclear Overhouser Effect (NOE) experiments to select conformers that provided the best match with the experimental data. In general the structures are selected by minimising a hybrid energy that connects a physical energy (e.g. from a forcefield) with experimental data

\begin{equation}
    \mathrm E_{\mathrm{hybrid}} = \omega_{\mathrm{data}}\cdot \mathrm E_{\mathrm{data}} + \mathrm E_{\mathrm{phys}}.
\end{equation}

However the methodology for evaluating agreement between structure and experimental data varies greatly, and is often somewhat arbitrary. Similarly the parameters and weights used for $\mathrm E_{\mathrm{data}}$ are often tweaked manually and optimal parameters seem to be based on trial and error.

The inferential structure determination (ISD) approach \citep{ISD,ISD:Science} uses a Bayesian formalism to handle these \emph{nuisance parameters}, such as the uncertainty and other model parameters, probabilistically as demonstrated by Olsson et al. \citep{Olsson} using a set of NOE restraints combined with a physical energy term.

This section introduces the ISD formalism for the Markov Chain Monte Carlo method simulations used to simulate both chemical shift assignment and protein structure.

\subsection{Probabilistic Framework}
The probability for event A given event B, $\mathrm P \left(A \mid B \right)$, is given by the chain rule

\begin{equation}
\mathrm P\left( A , B \right) = \mathrm P\left( A \mid B \right) \cdot \mathrm P\left( B \right),
\end{equation}
where $P\left( A , B \right)$ is the probability for both $A$ and $B$, which often is written as $P\left( A \cap B \right)$.

This, along with the equality $\mathrm P\left( A , B \right) = \mathrm P\left( B , A \right)$, leads directly to Bayes Theorem:

\begin{equation}
\mathrm P\left( A \mid B \right) = \frac{\mathrm P\left( B \mid A \right) \cdot \mathrm P\left( A \right)}{\mathrm P\left( B \right)}.
\end{equation}
\\

Using Bayes Theorem, we aim to find the most probable structure $X$, assignment $A$ and nuisance parameters $n$, given some experimental data $D$ and prior information $I$ (such as information used to generate the model describing the data, amino acid sequence etc.)
\begin{equation}
\mathrm P\left(X, A, n \mid D, I \right) = \frac{\mathrm P\left(D, I \mid X, A, n \right) \cdot \mathrm P\left(X, A, n \right)}{\mathrm P\left(D, I \right)}.
\end{equation}

Since only $X$, $A$ and $n$ are changed in Monte Carlo moves, terms not involving these doesn't need to be evaluated and can be disregarded, since the relative energy landscape is invariant of choice of normalisation constant.

\begin{align}
    \mathrm P\left(X, A, n \mid D, I \right)\nonumber
    &\propto {\mathrm P\left(D, I \mid X, A, n \right) \cdot \mathrm P\left(X, A, n \right)}\\ \nonumber
    &= {\mathrm P\left(D\mid I, X, A, n \right) \cdot \mathrm P\left(I \mid X, A, n \right) \cdot \mathrm P\left(X, A, n \right)}\\
    &= {\frac{\mathrm P\left(D\mid I, X, A, n \right) \cdot \mathrm P\left(X, A, n \mid I \right) \cdot \mathrm P \left( I \right) \cdot \mathrm P\left(X, A, n \right) }{\mathrm P\left(X, A, n \right)}}\\\nonumber
    &\propto {\mathrm P\left(D\mid I, X, A, n \right) \cdot \mathrm P\left(X, A, n \mid I \right)}\\\nonumber
    &= \mathrm P\left(D\mid I, X, A, n \right) \cdot \mathrm P\left( X \mid A, n, I \right) \cdot \mathrm P \left( A \mid n, I \right) \cdot \mathrm P\left( n \mid I\right)\\\nonumber
\end{align}

The prior distribution of $\mathrm P\left( n \mid I \right)$ is typically drawn from a log-normal distribution for purely positive parameters, and from a normal distribution if that's not the case. The argument being that these are the least biasing distributions according to the principle of maximum entropy \citep{Jaynes1,Jaynes2}. \\
$\mathrm P\left( X \mid A, n, I\right)$ is independent of $n$ and $A$. If a physical forcefield is used then the probability for a structure follows the usual Bolzmann distribution

\begin{equation}
    \mathrm P\left( X \mid I \right) = \frac{1}{Z} \cdot \exp\left(-\frac{\mathrm{E_{phys}}}{\mathrm{k_B \cdot T}}\right),
\end{equation}

Luckily we don't have to evaluate the partition function $Z$ since it appears as just a normalisation constant. $\mathrm P\left(X \mid I \right)$ can also be introduced as a generative probabilistic model (GPM) such as Torus-dbn \citep{TorusDBN} and Basilisk \citep{Basilisk} which replaces the physical term by a biased sampling of protein structure. These models are based on a large database of experimentally obtained structures backbone and side chain angles respectively.

For describing $P\left( D \mid I, X, A, n\right)$ the normal distribution is used because it's simple to work with mathematically and computationally. In addition, due to the Central Limit Theorem \citep{CLT}, the arithmetic mean of a large number of iterates of independent random variables will be approximately normal-distributed. A measured chemical shift $\delta_i$ will likely follow the distribution

\begin{equation}
\label{eq:true_gauss}
g(\delta_{i};\mu, \hat{\sigma}) = \frac{1}{ \hat{\sigma} \sqrt{2 \pi}}\, e^{-\frac{(\delta_{i} - \mu)^2}{2\hat{\sigma}^2}},
\end{equation}
with $\mu$ being the population mean (or "true" chemical shift) and $\hat{\sigma}$ being the standard deviation. 
The probability density of two independent measurements of a nuclei's chemical shift, $\delta_i$ and $\delta_j$ is then:

\begin{align}
f(\delta_{i}, \delta_{j};\sigma_i, \sigma_j)
&= \int_{-\infty}^{\infty}g(\delta_{i};\mu, \sigma_i) g(\delta_{j};\mu, \sigma_j) \pi \left(\mu\right) \mathrm d\mu \\
&\propto \left(\sigma_i^2+\sigma_j^2\right)^{-\frac{1}{2}}\exp\left(-\frac{\left(\delta_i-\delta_j\right)^2}{2\left(\sigma_i^2+\sigma_j^2\right)}\right)
\end{align}

Here $\mu$ has been integrated out using a uniform prior $\pi(\mu)$. Chemical shifts can be predicted using a forward model, such as {\tt SPARTA} \citep{SPARTA}, {\tt PROSHIFT} \citep{PROSHIFT}, {\tt SHIFTX} \citep{SHIFTX}, {\tt Camshift} \citep{Camshift} etc., which relates a structure to a set of chemical shifts. If $\delta_i$ is a predicted chemical shift value, then the corresponding standard deviation will be much larger than the experimental error. Upon taking the negative logarithm 

\begin{equation}\label{eq:gauss_pre}
    F_{pre}(\Delta_{ij};\sigma_i)
= \log\sigma_i + \frac{\Delta_{ij}^2}{2\sigma_i^2}
\end{equation}
with $\Delta_{ij} = \delta_i - \delta_j$. If both $\delta_i$ and $\delta_j$ are obtained from experiment and the same variance is assumed, then we get

\begin{equation}\label{eq:gauss_exp}
    F_{exp}(\Delta_{ij};\sigma)
= \log\sigma +\frac{\Delta_{ij}^2}{2\sigma^2}
\end{equation}
with $\sigma = 2\sigma_i=2\sigma_j$.\\

When more than two measurements of the same nuclei's chemical shift are used, things start to get more complex and some approximations are in order. For a predicted chemical shift $\delta_i$ and a set of experimentally obtained chemical shifts $\left\{\delta_j\right\}$, the following probability density is obtained

\begin{align}
\label{eq:difference_gauss}
f(\delta_i, &\left\{\delta_{j}\right\};\sigma_i, \sigma_j) 
= \int_{-\infty}^{\infty}g(\delta_{i};\mu, \sigma_i)\prod_j^N g(\delta_{j};\mu, \sigma_j) \pi \left(\mu\right) \mathrm d\mu \nonumber\\
&\appropto \frac{1}{\sigma_i}\exp\left(-\frac{\sum_j^N \left( \delta_i-\delta_j \right)^2}{2N\sigma_i^2}\right)\frac{1}{\sigma_j^{N-1}}\exp\left(-\frac{\sum_j^N \sum_{k>j}^N \left( \delta_j-\delta_k \right)^2}{2N\sigma_j^2}\right) \nonumber\\
&= \frac{1}{\sigma_i}\exp\left(-\frac{\chi_{pre}^2}{2N\sigma_i^2}\right)\frac{1}{\sigma_j^{N-1}}\exp\left(-\frac{\chi_{exp}^2}{2N\sigma_j^2}\right)
\end{align}
with $\chi_{pre}^2 = \sum_j^N \left( \delta_i-\delta_j \right)^2$ and $\chi_{exp}^2 = \sum_j^N \sum_{k>j}^N \left( \delta_j-\delta_k \right)^2$ where $k$ and $j$ refer to experimental chemical shifts.
The middle expression in \eqref{eq:difference_gauss} is obtained by tedious algebra with the only approximation used being $\sigma_i \gg \sigma_j$.\\

\eqref{eq:difference_gauss} can be approximated to the simpler form of \eqref{eq:gauss_pre} and \eqref{eq:gauss_exp} in order to simplify the calculations and reduce computational costs. Comparing these expressions, it is seen that if we make the approximation that every nuclei of the same type, have the same number of chemical shifts assigned to it, the negative logarithm of these expressions only differ by a normalisation factor. Using \eqref{eq:gauss_pre} to describe all interactions between the predicted chemical shift $\delta_i$ and the $N$ experimental ones $\left\{\delta_j\right\}$:

\begin{align}
    \sum_j^N F_{pre}(\Delta_{ij};\sigma_i)
&= \sum_j^N \left( \log\sigma_i + \frac{\Delta_{ij}^2}{2\sigma_i^2}\right) \nonumber \\
&= N\log\sigma_i + \frac{\chi_{pre}^2}{2\sigma_i^2}
\end{align}

Comparing this expression to \eqref{eq:difference_gauss} shows that the two equations differ by only a normalisation factor $\omega$:

\begin{align}
    \omega\left[ N\log\sigma_i + \frac{\chi_{pre}^2}{2\sigma_i^2}\right] &= \log\sigma_i + \frac{\chi_{pre}^2}{2N\sigma_i^2} \\
                                                                  \omega &= \frac{1}{N}
\end{align}

Similarly, \eqref{eq:gauss_exp} can be used to describe all unique pairings of the experimental chemical shifts. For $N$ chemical shifts, there will be a total of $N\left(N-1\right)/2$ unique pairings (given by $\sum_j^N \sum_{k > j}^N$), resulting in:

\begin{align}
    \sum_j^N \sum_{k > j}^N  F_{exp}(\Delta_{jk};\sigma_j)
&= \sum_j^N \sum_{k > j}^N \left(\log\sigma_j +\frac{\Delta_{jk}^2}{4\sigma_j^2}\right) \nonumber \\
&= \frac{N\left(N-1\right)}{2}\log\sigma_j +\frac{\chi_{exp}^2}{4\sigma_j^2}
\end{align}
where constant terms have been neglected.
Note the factor of 4 in the denominator of the right-most term instead of a factor of 2, due to not replacing $\sigma_j$ with $\sigma$. Comparing with \eqref{eq:difference_gauss} to find the normalisation factor:

\begin{align}
    \omega\left[ \frac{N\left(N-1\right)}{2}\log\sigma_j + \frac{\chi_{exp}^2}{4\sigma_j^2}\right] &= \left(N-1\right) \log\sigma_i + \frac{\chi_{exp}^2}{2N\sigma_i^2} \\
                                                                  \omega &= \frac{2}{N}
\end{align}

To summarise, considering only the disagreement between predicted and assigned chemical shifts, with a total of $N_j$ experimentally measured chemical shifts assigned to nuclei of the same type for $j \in \left\{ C^\alpha, H, N, C, C^\beta \right\}$,

\begin{align}
    \mathrm P_{pre}\left(D \mid X, A, \left\{ \sigma_{pre,j} \right\}, I\right)
    & \propto \prod_j \prod_i^{N_j} \left[ \frac{1}{\sigma_{pre,j}} \exp \left( -\frac{\Delta_{ij}^2}{2\sigma_{pre,j}^2}\right)\right]^{\omega_{pre,j}}\\
    & = \prod_j \left(\sigma_{pre,j}\right)^{-N_j\omega_{pre,j}} \exp \left( -\frac{\chi_{pre,j}^2\omega_{pre,j}}{2\sigma_{pre, j}^2}\right)
\end{align}
where $\Delta_{ijk}$ is the difference between chemical shift $i$ and the predicted chemical shift $k$ for nuclei type $j$, $\chi_{pre,j}^2 = \sum_i^{N_j} \Delta_{ijk}^2$ and $\omega_{pre,j}$ is the weight for nuclei type $j$. Its exact weight can estimated from the number of contributions to $\chi_{pre,j}^2$ in the simulation.\\
Likewise the disagreement between chemical shifts from different experiments assigned to the same atom is treated in the same manner, but with separate nuisance parameters $\left\{\sigma_{exp,j} \right\}$.
\begin{equation}
    \mathrm P_{exp}\left(D \mid A, \left\{\sigma_{exp,j} \right\}, I \right) 
\propto \prod_j \left(\sigma_{exp,j}\right)^{-m_j\omega_{exp,j}} \exp \left( -\frac{\chi_{exp,j}^2\omega_{exp,j}}{2\sigma_{exp,j}^2}\right)
\end{equation}
with $\chi_{exp,j}^2$ containing a total of $m_j$ unique chemical shifts differences. \\
\\
$\mathrm P \left( A \mid n, I \right)$ basically describes the probability density for having $N_j$ chemical shifts assigned. Since a complete one to one assignment of all peaks usually is impossible, a model describing whether an assignment is better or worse than having no assignment at all is needed. Currently every "missing" contribution to $\chi_{pre,j}^2$ is replaced by a chemical shift difference of $3\sigma_{pre,j}$. The effect of this is that assignment will be favoured if the chemical shift differences are lower than $3\sigma_{pre, j}$, and unassignment will be favoured if it is not. Likewise for $\chi_{exp,j}^2$, missing contributions is replaced by a difference of $4\sigma_{exp,j}$. These exact values were chosen since they seem to perform the best.\\

Putting it all together when a physical force field is used, the probability distribution we aim to simulate will be:

\begin{multline} \label{eq:hybrid_energy}
\mathrm P\left(X, A, n \mid D, I \right)\propto \\ 
\exp\left(-\frac{\mathrm{E_{phys}}}{\mathrm{k_B \cdot T}}\right)\prod_j \frac{\sigma_{pre,j}^{-N_j\omega_{pre,j}}}{\sigma_{exp,j}^{m_j\omega_{exp,j}}} \exp \left( -\frac{\chi_{pre,j}^2}{2\sigma_{pre,j}^2}-\frac{\chi_{exp,j}^2}{2\sigma_{exp,j}^2}\right) \cdot \mathrm P \left( n \mid I\right)
\end{multline}
where $\mathrm P \left( n \mid I \right)$ will be removed as a bias in the acceptance rate (See Section~\ref{section:n_moves}). The associated hybrid energy is

\begin{multline}
    \mathrm{E_{hybrid}} = E_{phys} \\
    +\mathrm{k_B T} \sum_j \left[\omega_{pre, j}\left(N_j \log\sigma_{pre,j} + \frac{\chi_{pre,j}^2}{2\sigma_{pre,j}^2}\right) + \omega_{exp,j}\left(m_j \log\sigma_{exp,j} +\frac{\chi_{exp,j}^2}{2 \sigma_{exp,j}^2}\right)\right]
\end{multline}

Since the structure $X$, assignment $A$ and parameters $n$ are all treated as variables, Monte Carlo moves are needed for each of these 'dimensions' of the sampling space as described in the next section.

\section{Computational Details}
{\tt Phaistos} is a software framework for Markov chain Monte Carlo sampling for simulation, prediction, and inference of protein structure \citep{Phaistos}. A large range of Monte Carlo moves is implemented for structure inference with selected physical force fields, and so is state of the art Monte Carlo methods and the forward model Camshift. In addition to this the probabilistic framework makes it easy to implement and treat empirical inferred models of experimental data together with physical forcefields in a rigid probabilistic fashion, which has been done previously for NOE's \citep{Olsson}.

\subsection{Markov Chain Monte Carlo}
Markov Chain Monte Carlo (MCMC) algorithms sample from probability distributions in the steady state, and are desirable to use when the distribution isn't easily expressible analytically. The probability distribution of a set of variables $\left\{x\right\}$ can be approximated by this method, given that a function $f(\left\{x\right\})$ that's proportional to the real distribution is known. \\
\\
The most common MCMC method is the Metropolis-Hastings algorithm \citep{Metropolis}. Given the most recent sampled state $x_t$, a new state $x'$ is proposed with a probability density that adhere to detailed balance

\begin{equation}\label{eq:detailed_balance}
\mathrm P\left(x_t\right) \mathrm P\left( x_t \rightarrow x'\right) = \mathrm P\left(x'\right) \mathrm P\left( x' \rightarrow x_t\right)
\end{equation}
which in turn ensures that samples correspond to the steady state. If the probability for this state is greater than the previous state, the proposed new state is accepted and $x_{t+1} = x'$. If the probability is lower, the Metropolis-Hastings acceptance criteria of the proposed state is given by

\begin{equation}
    \mathrm{P_{acc}} = \min \left( 1, \frac{f(x')}{f(x_t)}\right)
\end{equation}

\begin{figure}
    \centering
    \includegraphics[width=0.25\textwidth]{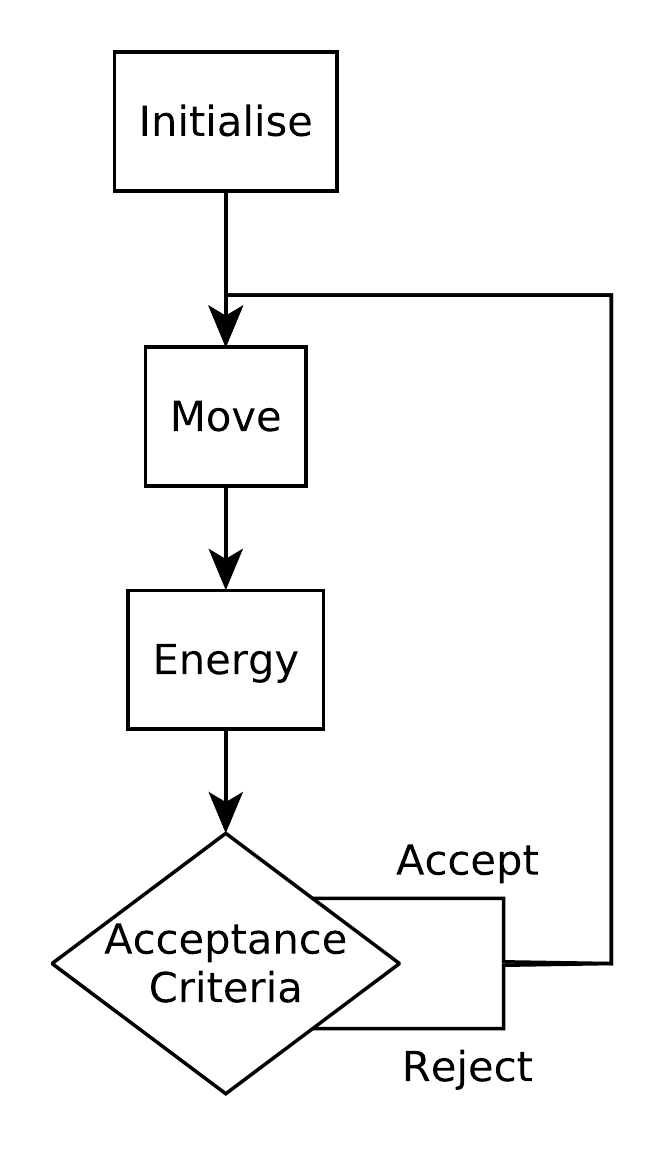}
    \caption{Flowchart showing the steps of the Metropolis-Hastings algorithm.}
    \label{fig:metropolis-hastings}
\end{figure}

If the state is rejected the system will return to the previous state $x_{t+1} = x_t$. The Metropolis-Hastings algorithm is shown schematically in Figure~\ref{fig:metropolis-hastings}

Other more advanced MCMC methods is implemented in {\tt Phaistos}, but all simulations run so far have been using the Metropolis-Hastings method. However since all implemented Monte Carlo moves in {\tt Chemshift} uphold detailed balance, other methods can easily be used as well.

\subsection{{\tt Chemshift} implementation in {\tt Phaistos}}
The Monte Carlo method requires both evaluation of energy and Monte Carlo moves that propose new values for the sampled parameters. The hybrid energy used is described in Section~\ref{section:Theory} and the Monte Carlo moves used for assignment is presented here.\\

Each spectrum of the types, HSQC, HNCA, HNcoCA, HNcoCACB, HNCACB, HNCO and HNcaCO that are available, is parsed from their input files where each peak is split into the chemical shifts according to the originating nuclei, as shown below: \\

$\left[ C^{\alpha}_{i-1}, H_{i-1}, N_{i-1}, C_{i-1}, C^{\beta}_{i-1},C^{\alpha}_{i}, H_{i}, N_{i}, C_{i}, C^{\beta}_{i}\right]$
\\
\\
Unused sites in these constructed peak-lists are given a {\tt NAN} value to be easily recognisable. If the peak is assigned to a specific spin system in the input file the same assignment is used in the module. All spin systems that have not been assigned a peak is assigned a list with only {\tt NAN} values.
This results in an array initially the same length of the protein. All the unassigned peaks is placed at the back of this array in an "unassigned" region, where the energy isn't evaluated. This procedure is repeated for all the spectra available.
The spectra HNCA, HNcoCACB, HNCACB and HNcaCO contain peaks from more than one backbone spin system and an array is created for each spin system type. As an example HNCA is split into an inter-peak and intra-peak array. For HNCA and HNcaCO, unassigned peaks are placed randomly in the unassigned region of the inter and intra array, and for HNcoCACB the largest carbon chemical shifts is attributed to $C^\alpha$. For HNCACB, which contains four peaks per residue, peaks from $C^{\alpha}$ and $C^{\beta}$ are assumed to be of opposite phase, and the nuclei type can be uniquely identified. Whether a peak is placed in the nuclei specific inter or intra peak is random.

\subsubsection{Monte Carlo Nuisance Parameter Moves}
\label{section:n_moves}
$\sigma$ describes the always positive standard deviation, so the log-normal distribution is well suited to propose new values for this. However by imposing this distribution for the data, a small bias will be introduced in the acceptance criteria, since

\begin{equation}
    \mathrm{P_{acc}} \propto \frac{\mathrm P \left( \sigma' \mid I\right)}{\mathrm P \left( \sigma \mid I \right)}
\end{equation}

From detailed balance \eqref{eq:detailed_balance} this bias is removed by multiplying with

\begin{equation}
    \frac{\mathrm P \left(\sigma' \rightarrow \sigma \right) }{\mathrm P \left(\sigma \rightarrow \sigma' \right)}
\end{equation}
whenever a move in the nuisance parameter space is made.

The {\tt update_sigma} move make changes to a single element in $\left\{\sigma_{pre,j}\right\}$ or $\left\{\sigma_{exp,j}\right\}$. Specifically this is done by drawing a factor $x$ from a log-normal distribution with parameters $\mu = 0$ and $\sigma_{\sigma} = 1$.

\begin{equation}
\mathrm P\left(x\right) \propto \frac{1}{x}\exp\left(\frac{\log^2x}{2}\right)
\end{equation}

The proposed new value $\sigma'$ for the standard deviation is
\begin{equation} \label{eq:sigma_sampling}
    \sigma' = \sigma \cdot x \quad \Leftrightarrow \quad \sigma = x^{-1} \sigma'.
\end{equation}

The corresponding bias that needs to be included in the acceptance criteria for the move is then

\begin{align}
    \frac{\mathrm P \left(\sigma' \rightarrow \sigma \right) }{\mathrm P \left(\sigma \rightarrow \sigma' \right)} &= \frac{\mathrm P\left(x^{-1}\right)}{\mathrm P\left(x\right)} \\
    &= \frac{\left(x^{-1}\right)^{-1} \exp \left(- \frac{\left(\log{x^{-1}}\right)^2}{2}\right)}{\left(x\right)^{-1} \exp \left(- \frac{\left(\log{x}\right)^2}{2}\right)}\\
    &= x^{2}
\end{align}

\subsubsection{Monte Carlo Assignment Moves}
To ensure that a specific assignment can be reached (at least in theory) in the simulation, it's important to cover the entire assignment-space. This is done by the following five moves:\\

{\tt move_single} picks an array at random and interchanges two peaks in this array, providing the means to switch assignments, unassign previously assigned peaks and vice versa. \\

{\tt move_HNCA} works the same as above, but instead of interchanging two peaks in the same array, a peak from the inter HNCA array is interchanged with a peak in the intra HNCA array, followed by a reclassification of the chemical shift assigned from $C^\alpha_{i-1}$ to $C^\alpha_i$ and vice versa.\\

{\tt move_HNcoCACB and move_HNcaCO} are similar to the above, just with the arrays made from the HNcoCACB and HNcaCO spectra respectively.\\

{\tt move_CA_HNCACB and move_CB_HNCACB} moves between the spin systems $C^\alpha_{i-1}$ and $C^\alpha_i$ and likewise for $C^\beta$. Changing a $C^\alpha$ assignment to a $C^\beta$ assignment is not possible, since it is assumed that these are always distinguishable by their phase.\\

During both a manual and simulated assignment, a ladder of spin systems connected through their intra and inter peaks can usually be constructed, where the created sequence of peaks matches very well. If these ladders are incorrectly assigned, it will be very difficult to reassign them with moves that only interchange two peaks at a time, due to a low acceptance rate. Because of this a set of moves that can reassign parts of or whole ladders is implemented.

These moves are carried out in two functions, {\tt move_base} and {\tt move_peak_blocks}, with several Monte Carlo moves utilising these with different parameters.

{\tt move_base} is used by a range of Monte Carlo moves to reassign 1 to $N$ adjacent peaks from 1 to $M$ different spin system arrays simultaneously, but doesn't change which array each peak is placed in.
The number of arrays involved in the move depends entirely on arbitrary chosen weights. These weights will only affect how fast the simulation reaches convergence etc. and not the energy landscape as such. Because of this no rigorous optimisation of these parameters has been done.
The probability for selecting a specific number of adjacent peaks is arbitrary as well, but smaller numbers are more probable than higher numbers, and the probability approximately follows an exponential decay with increasing ladder size.

In the initialisation steps of the module, an array is generated with every possible placement for ladders of size 1 to $N$ which make $N$ equal to the size of the largest segment in the protein with no Prolines. The placement of Glycines in the protein is noted in this array as well to make sure no $C^\beta$ chemical shift are assigned there.
The move itself, given a number of adjacent peaks to move in a number of spin system arrays, is often non problematic and two peak "blocks" swap assignments. If a Glycine is present in one of these protein segments, any peak with a $C^\beta$ chemical shift that would wrongly be assigned to the Glycine is instead moved to the unassigned region.\\
When a ladder is moved a smaller distance than the length of the ladder itself, the problem arises that the starting assignment of the ladder overlaps with the destination of the ladder. An example is shown below, with $i_n$ being peaks that are to be moved to sites $j_n$.
\begin{equation}
\left[\;i_0\;,\;i_1\;,\; i_2\;,\; i_3\;,\; i_4\; /\; j_0\;,\; i_5\; /\; j_1\;,\; j_2\;,\; j_3\;,\; j_4\;,\; j_5\;\right]\nonumber
\end{equation}

For this situation special care is needed in order to conserve as much integrity of the moved ladders as possible.To achieve this one full ladder is selected at random from the two overlapping ones, and this ladder will be moved as it is, with the resulting assignment shown below\\
\begin{equation}
\left[\; j_2\;,\; j_3\;,\; j_4\;,\; j_5\;,\;i_0\;,\;i_1\;,\; i_2\;,\; i_3\;,\; i_4\; /\; j_0\;,\; i_5\; /\; j_1\;\right]\nonumber
\end{equation}

{\tt move_peak_blocks} is of similar construct, but interchanges two ladders from different spin system arrays, originating from the same experiment.\\

\begin{figure}    \centering
    \includegraphics[width=1\textwidth]{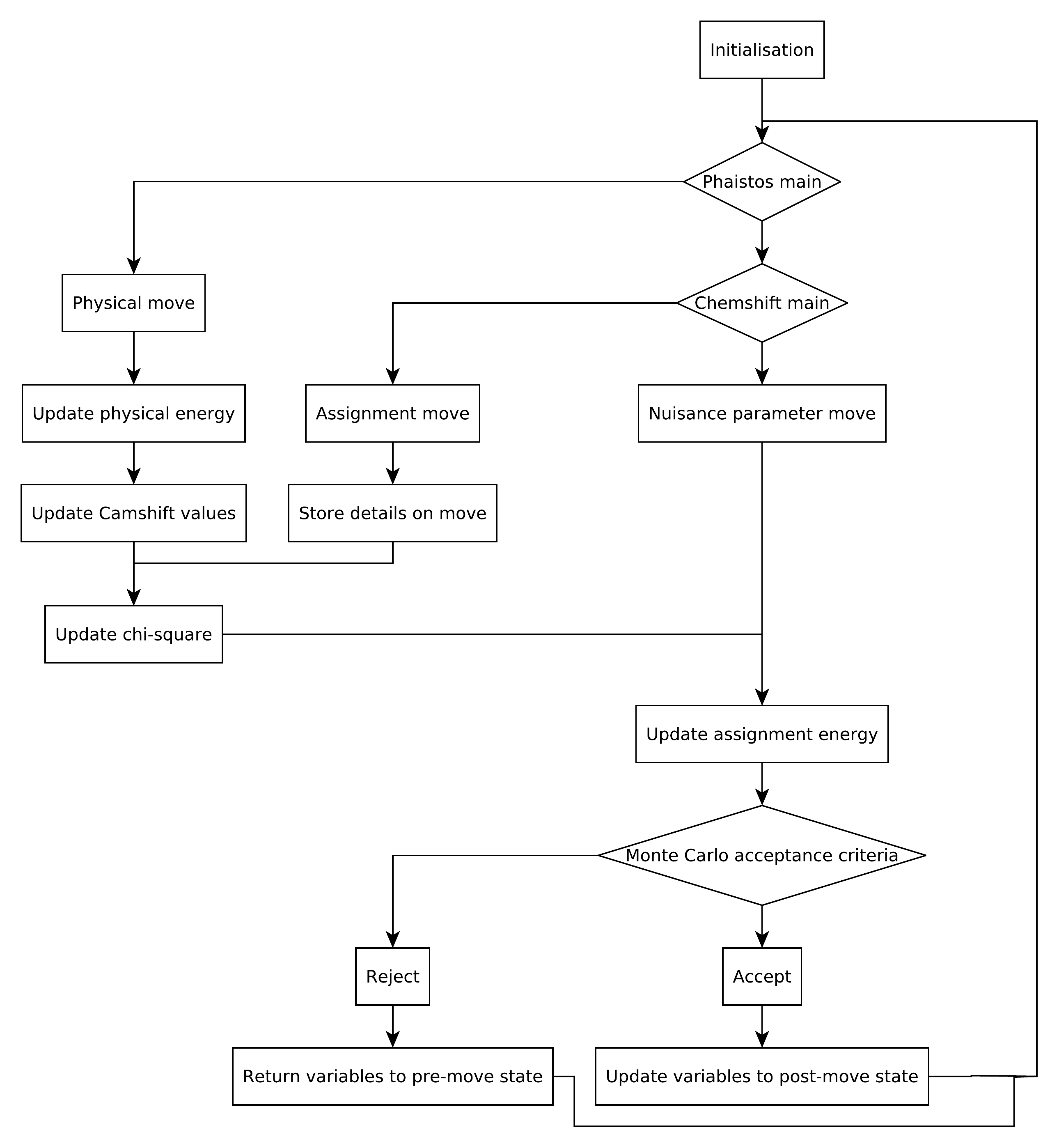}
    \caption{Flowchart showing the general strategy in a Monte Carlo simulation in {\tt Phaistos} with the {\tt Chemshift} module. Details in the text}
    \label{fig:flowchart}
\end{figure}

Figure~\ref{fig:flowchart} shows a simplified flowchart of a Monte Carlo simulation with {\tt Chemshift}.

\subsubsection{Cashing}
The computational aspect of this project represents around 90\% of the work done. Other than on implementation and development of the different aspects of the program, a considerate amount of time have been used on increasing the speed of the calculations.\\
In the initialisation part of the program, starting guess values is set for the nuisance parameters, the Camshift predictions are created and the sum of all possible chemical shift differences ($\chi^2$) is calculated. This last step takes a very long time and would be a major bottleneck if it were to be run after each move. To reduce the time used, two functions, {\tt initialise_chi_sq_details} and {\tt initialise_chi_sq_partial} are employed. \\
The first function scans through each spin system array and notes which chemical shift types the array contains, and stores all the possible permutations of chemical shift differences that can arise. That is it won't try to check the $C^\beta$ differences between HNCA and HNCACB peaks, since the $C^\beta$ values will always be {\tt NAN} in the HNCA as well as the $C^\alpha$ spin system arrays of HNCACB.
The second function stores every contribution to $\chi^2$ separately instead of just storing the sum. In every iteration, information about what move is used, which spin system array change and which peaks are moved is stored, making it possible to both reverse the move made if it is rejected, instead of having to save and copy the complete assignment every iteration, but also to use the information from {\tt initialise_chi_sq_partial} to only calculate the contributions that are changed. \\

Knowing which spin systems the changed peaks were and became assigned to cuts down calculation cost dramatically. However further reducing the number of calculations done, to only include the spin system arrays that were moved in is a bit more complicated.
When only changes are made in one spin system array, only the chemical shift differences between this array and all the others need to be updated (disregarding Camshift predicted chemical shifts for the moment, as calculation of these is trivial). If changes are made in all the spin system arrays, all terms have to be updated. However in between these extremes the computational part is a bit more complex, even though only the differences between just the changed spin system arrays, and the difference between the changed and the non-changed arrays need to be calculated. \\
Because of this extra (but not easily recognised) computational cost, this procedure is only done on $H$ and $N$ chemical shifts, while all possible differences are calculated for the rest of the nuclei. The argument for doing it this way is that, given the spectra HSQC, HNCA, HNCACB, HNcoCACB, HNCO and HNcaCO, there will be 66 possible differences to be calculated for $H$ and $N$ each, 10 for $C^\alpha$ and 3 for $C$ and $C^\beta$ each. So carbon differences is only about 10\% of all the contributions, and it didn't seem like any noticeable benefit in computational cost would be gained. \\

During these simulations, the assignment itself, as well as the nuisance parameters, $\chi^2$ and the list containing every contribution to $\chi^2$ need to be able to be returned to the previous state if the move is rejected. Just keeping and updating copies of these after every iteration would be a major bottleneck, so if a move is rejected, the moves are written such that the previous state can be regained by using the same move type, with the same parameters. The list with $\chi^2$ contributions, could be updated in a similar fashion, but a faster way is to keep a copy of the list, and instead of copying the full list every iteration, use the stored move information to only copy the terms that may have changed. \\

Currently an average of 2.6 billion assignment moves per day can be done on the 101 residue protein S6 on a single 3.0 GHz Xeon core, with around 10\% of the time spent being overhead from {\tt Phaistos} itself. In comparison around 2.8 million Camshift predictions can be done per day, and further improvements to the speed of the program have been halted until it becomes a bottleneck in the protein folding process.

\section{Results}
A range of simulations have been run on Ribosomal Protein S6, for the purpose of testing the accuracy and breaking points of the assignment model, given a crystal structure. S6 was chosen for the simple reason that it's the only protein where a manual assignment, Autoassign assignment and FLYA assignment for individual peaks have been available to us. In these simulations no changes were being made to the structure.  \\

\begin{figure}
    \centering
      \includegraphics[width=0.4\textwidth]{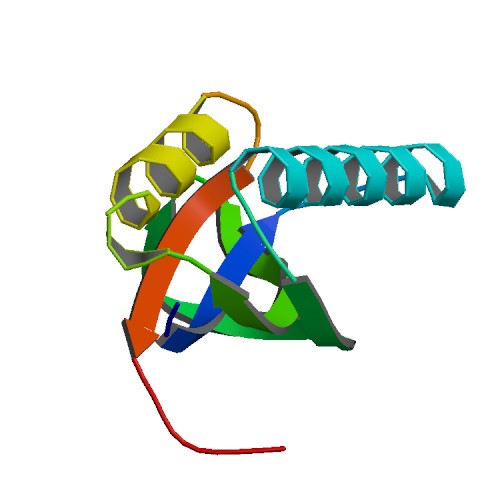}
      \caption{The 101 residue Ribosomal Protein S6 (PDB:1LOU)}
      \label{S6}
\end{figure}

Using HSQC, HNCA, HNCO, HNcaCO, HNCACB and HNcoCACB spectra, the 101 residue protein could theoretically be assigned 1327 peaks, with 950 peaks being assigned in the manual assignment.\\

\begin{figure}
    \centering
      \includegraphics[width=0.9\textwidth]{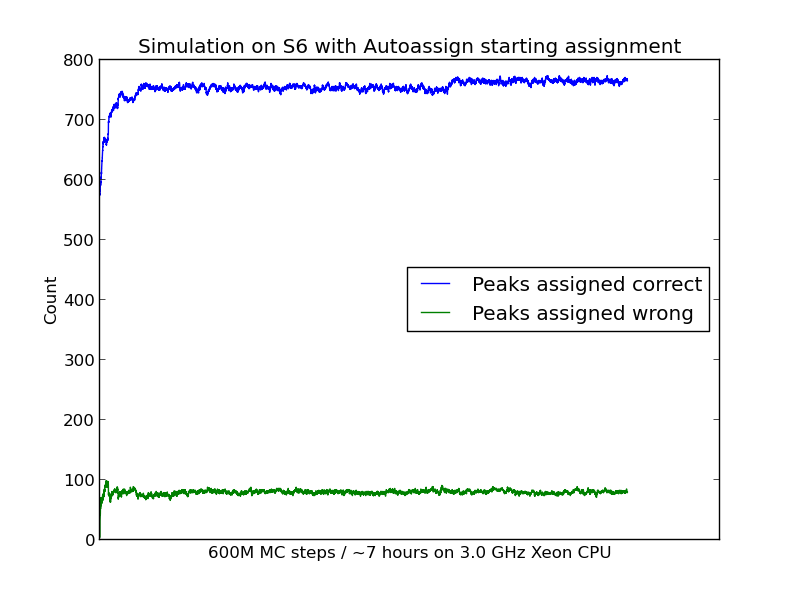}
      \caption{Simulation on S6 with assignment and nuisance parameter moves, with the initial assignment being done by Autoassign. Peaks were deemed correct if all chemical shifts of the peak were within the tolerance region of 0.03 ppm for Hydrogen and 0.4 ppm for the heavy nuclei, compared to the manual assignment}
      \label{aa_MC}
\end{figure}

The agreement between the manual assignment and assignments obtained via the simulations was investigated, for four different starting assignments. The manual assignment, the FLYA assignment, the Autoassign assignment and finally starting with a random assignment.\\
Figure~\ref{aa_MC} shows the number of peaks correctly assigned as the simulation progresses. A peak is considered correctly assigned if all chemical shifts of the peak lies within 0.03 ppm for hydrogen and 0.4 ppm for the heavy nuclei compared to the manual assignment, which is the same criteria used in the FLYA paper.

The assignment by Autoassign agrees with the manual assignment for 575 peaks initially. As the simulation progresses, this number rises to around 770 while the number of peak assignments that disagrees with the manual assignment rose from 5 to around 80. The fact that a large number of chemical shifts is being incorrectly assigned isn't as troublesome as it would be for a deterministic assignment, since each point in Figure~\ref{aa_MC} represents a snapshot of the assignment at a particular time. If the most probable assignment of a peak was taken from a histogram of all the assignment snapshots, the number of incorrect assignments would quite possibly be lower than what appears from the figure. However this trend would also be likely to be observed if the energy function used to describe the experimental data is of poor quality. \\

\begin{figure}
    \centering
      \includegraphics[width=0.9\textwidth]{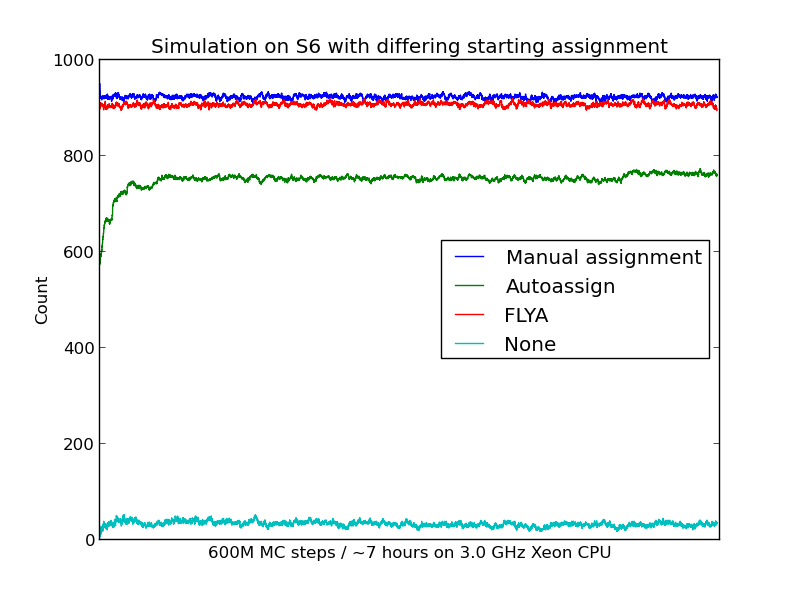}
      \caption{Number of correcly assigned peaks with initial assignment done manually, by FLYA, by Autoassign and no initial assignment at all.}
      \label{S6_start}
\end{figure}

Figure~\ref{S6_start} shows the agreement of the simulation with the manual assignment, starting from different initial assignments. When starting from the manual assignment, the agreement went down as expected from 950 initially to around 924 peaks on average, with no incorrect assignments. FLYA experienced little change, going from 908 initially to 904 correct on average, with the number of incorrectly assigned peaks dropping from 18 initially to 14 on average. \\
When a random initial assignment was given, the simulation was quickly stuck in a local minimum with very poor agreement on especially $H$ and $N$ nuclei chemical shifts, which could either be a sampling problem, or due to a poor model description.\\

\begin{figure}
    \centering
      \includegraphics[width=0.9\textwidth]{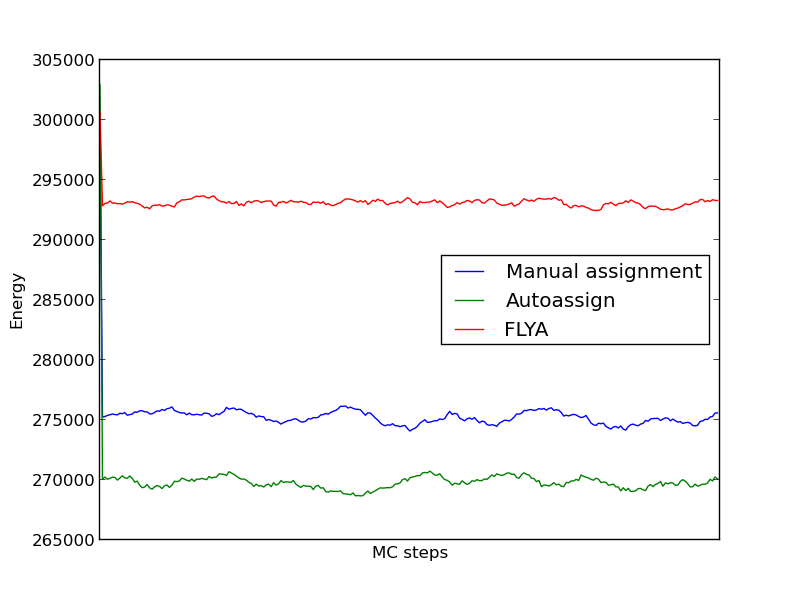}
      \caption{Energies of three simulations on S6, with three different starting assignments, consisting of only nuisance parameter sampling.}
      \label{S6_sigma_sampling}
\end{figure}

Investigating the energies of the different starting assignments, using only nuisance parameter moves (no changes being made to the assignment), the energies is expected to follow $E_{autoassign} > E_{FLYA} >= E_{manual}$, based on the correctness of the assignments. Surprisingly the energies were found as following $E_{FLYA} > E_{manual} > E_{Autoassign}$ as shown in Figure~\ref{S6_sigma_sampling}.

That Autoassign is lowest in energy strongly suggests that the model for describing unassigned chemical shifts needs to be improved. However the difference between the manual assignment and the FLYA assignment cannot be explained simply by this, since they should be very similar. Therefore it is very clear that improvements in general of the energy function is critical for improving upon the current assignment capabilities of the module. %

\section{Future Work}
\label{section:Improvements}
The {\tt Chemshift} module is as previously stated a work in progress, and in terms of module functionality, a number of improvements is planned. The most important being model improvements. In the following, planned improvements to the model, that have yet to be implemented, is presented.

\subsection{Referencing Errors}
From the simulations on S6, it is clear that improvements to the energy function needs to be made.

\begin{figure}[htb]
  \centering
  \subfloat[]{%
    \includegraphics[width=.52\textwidth]{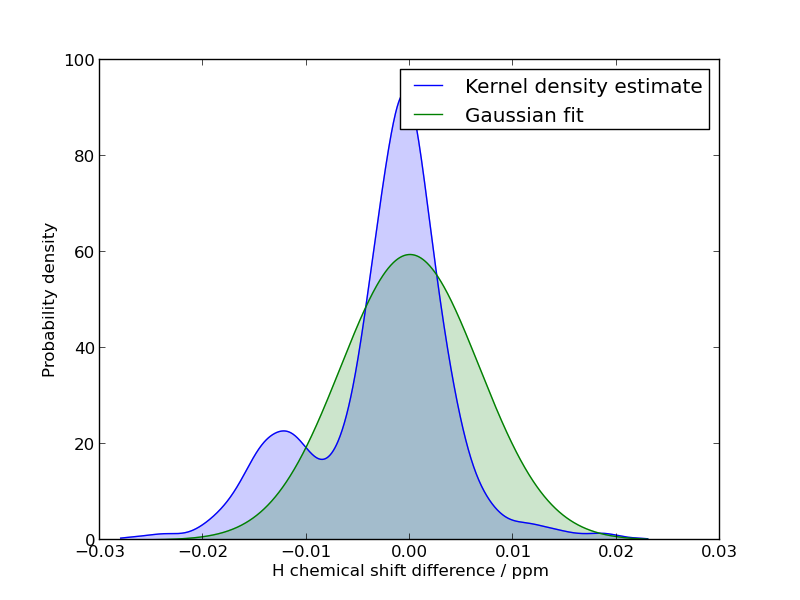}}\hspace{-20pt}
  \subfloat[]{%
    \includegraphics[width=.52\textwidth]{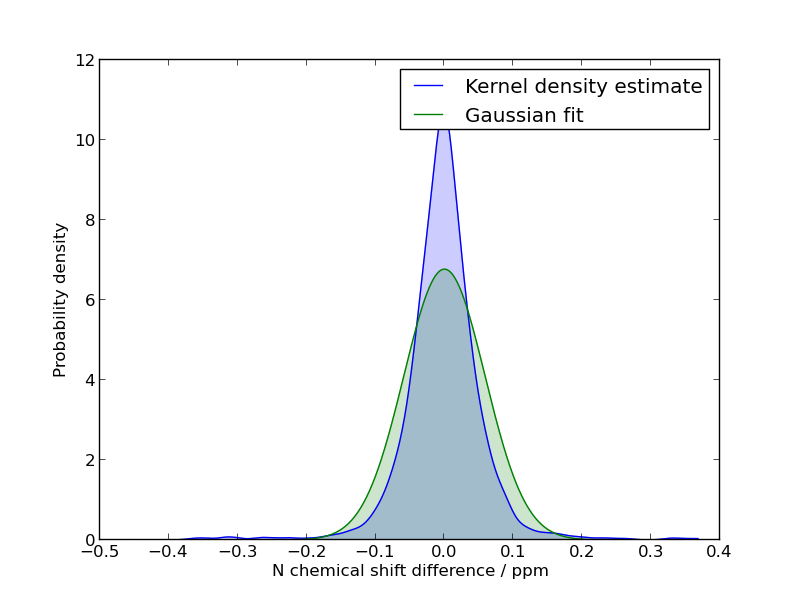}}\\
  \subfloat[]{%
    \includegraphics[width=.52\textwidth]{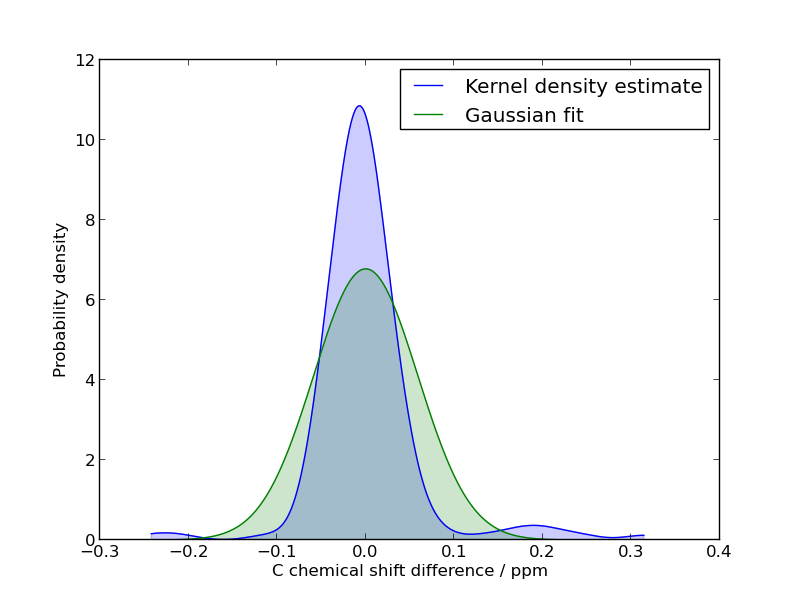}}\hspace{-20pt}
  \subfloat[]{%
    \includegraphics[width=.52\textwidth]{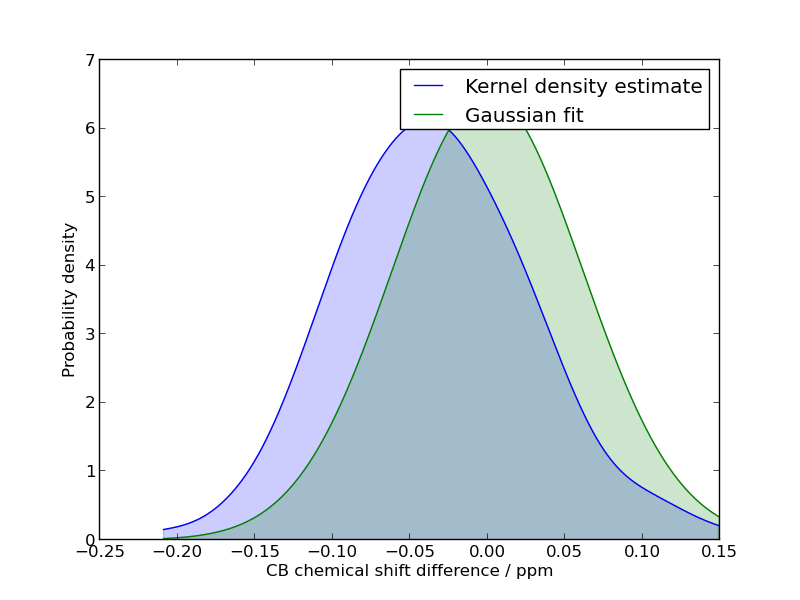}}\\
  \caption{Differences between chemical shifts assigned to the same nuclei from S6. Blue graph show the Kernel Density Estimate for the data, while green shows the best fit with a normal distribution}
  \label{fig:s6_diffs}
\end{figure}

As shown in Figure~\ref{fig:s6_diffs} the current model describes actual data from the protein S6 somewhat poorly in some cases.

The description of $H$ chemical shifts is especially poor and a likely cause of this is small perturbation differences to the reference nuclear shielding. In other words, the spectra used isn't properly aligned.

This alignment correction would correspond to a small correction to each chemical shift, depending on which spectra it originates from. The chemical shift difference for hydrogen from HSQC and HNCO would be \\$\left(\left(\delta_{HSQC}+\gamma_{HSQC}\right)-\left(\delta_{HNCO}+\gamma_{HNCO}\right)\right)$ instead of just $\left(\delta_{HSQC}-\delta_{HNCO}\right)$, with $\gamma_i$ representing the alignment offset of spectra $i$. These values of $\gamma_i$ could be treated as a nuisance parameter, with sampling done from a normal distribution.\\

\begin{figure}
  \centering
    \includegraphics[width=.6\textwidth]{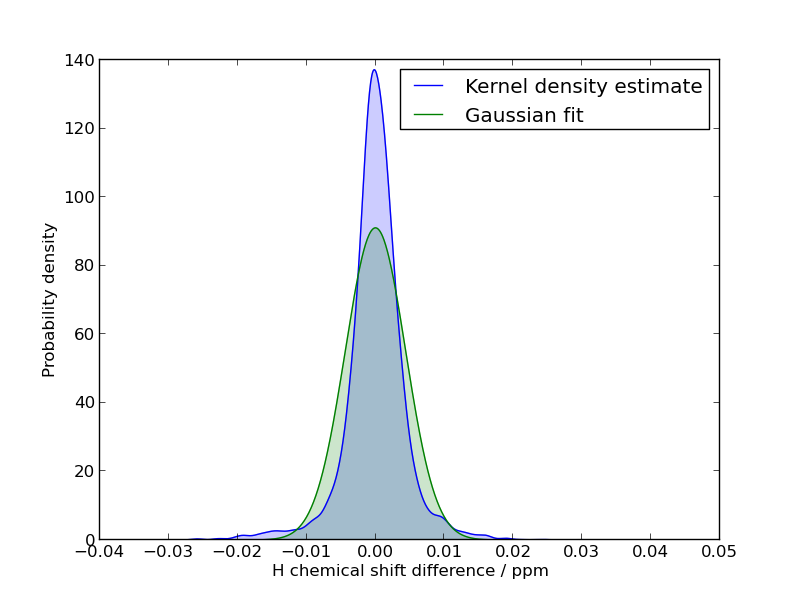}
    \caption{Differences between chemical shifts assigned to the same H nuclei, after alignment.}
    \label{fig:s6_corrected_diffs}
\end{figure}

Correcting the S6 spectra, with values of $\gamma_i$ that maximises the model likelihood, the hydrogen differences obtained follow the simple Gaussian model much closer as seen in Figure~\ref{fig:s6_corrected_diffs}.

\subsubsection{Model Validation}
When comparing different models, just a visual determination of the best model is prone to be erroneous. In addition adding parameters to be fitted will always improve a model, but might end up causing a low predictive validity due to over-fitting.\\
To determine if the increase in goodness of fit outweighs the increase in complexity of the model (ignoring increased computational cost for the moment), Aikake's Information Criterion (AIC) can be used \citep{AIC}. AIC is a measure of the relative quality of a given model, and can be used for model selection, where the model with the minimum AIC value is prefered. \\
The AICc is an improved version of the AIC that includes corrections for finite sample size, and should in general always be used instead of the AIC \citep{AICc}. The AICc is given by:

\begin{equation}
    AICc = 2k-2\log\left(L\right)+\frac{2k\left(k+1\right)}{n-k-1},
\end{equation}
with $k$ being the number of parameters in the model, $n$ being the sample size and $L$ being the maximum value of the likelihood function (the joint density function for all observations) for the estimated model.\\

For the Gaussian model for $H$ differences with no alignment, the only parameter is the standard deviation. Maximising the likelihood of the S6 data yields an AICc value of -32060.97. Including alignment adds 5 new parameters when 6 spectra is used and yields an AICc value of -35858.28, which suggests that the improvement in goodness of fit is worth the information lost by increasing the number of parameters.

\subsection{Peak Intensities}
\begin{figure}
  \centering
    \includegraphics[width=.6\textwidth]{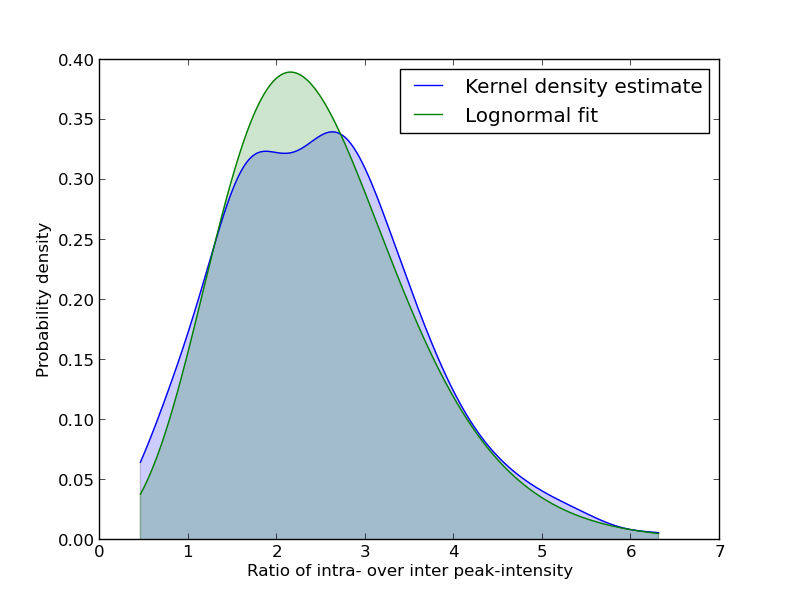}
    \caption{Ratio of all intra- over inter peak intensities for carbon atoms in the S6 HNCACB, HNcaCO and HNCA spectra}
    \label{fig:peak_ratio}
\end{figure}

In experiments containing both inter and intra peaks, the intra peak has a higher intensity on average than the inter peak, with an average ratio of around 1.5 having been reported \citep{HNCA}. But since there's a large variance in this ratio, and ratio's less than 1 often is observed, these intensities are often ignored by experimentalists. But for a probabilistic model, it should provide valuable information. \\
Figure~\ref{fig:peak_ratio} shows these peak ratios for S6. Since the peak ratios approximately follow a log-normal distribution, it should be easy to implement this as an energy-term as well.

Of course the model selection will need to be validated on more than a single protein. Other model improvements that need to be investigated include describing data with different standard deviations depending on which spectrum it is from, using a function family other than the normal distribution, include possible correlation between different atom types and improving how unassigned chemical shifts is treated.

\section{Summary and Outlook}

This thesis presents the current state of a new method for including experimental NMR data in protein structure determination, and the method has been implemented in the protein structure inference program {\tt Phaistos}. The most noteworthy features is that 1) no peaks in the experimental spectra is discarded, providing more information about the structure than a regular deterministic assignment. 2) The assignment can change during protein folding, possibly giving a better description of the protein dynamics and reducing the effect of assignment errors. 3) The weight of experimental data relative to physical energy terms, is decided probabilistically instead of relying on arbitrary manual weights.\\

By running simulations on the 101 residue Ribosomal Protein S6, some improvement to a partial assignment done by the program Autoassign has been made. By analysing the energies of assignments of differing qualities, it is clear that improvements need to be made to the proposed model. Improvements such as sampling the referencing errors between spectra and including additional energy terms related to peak intensities has been proposed based on statistical observations.\\

Due to time restraints a proper validation of the method, by successfully folding a range of proteins, using unassigned chemical shift experiments, have yet to be done. However the entire framework for doing so has been created, and doing this is the intent of the project. \\
Assuming that validation of the method is possible, the generated framework can easily be used to include assignment of protein side chain nuclei or to assign NOE's at the same time as the chemical shifts. Furthermore histograms over the assignment of each peak could be generated to assist manual assignments.\\
Over the next several months, work will continue on the {\tt Chemshift} module, which will eventually be included in the official {\tt Phaistos} release.

\clearpage
\newpage
\section{Appendix}
\begin{figure}[htb]
  \centering
  \subfloat[]{%
    \includegraphics[width=.52\textwidth]{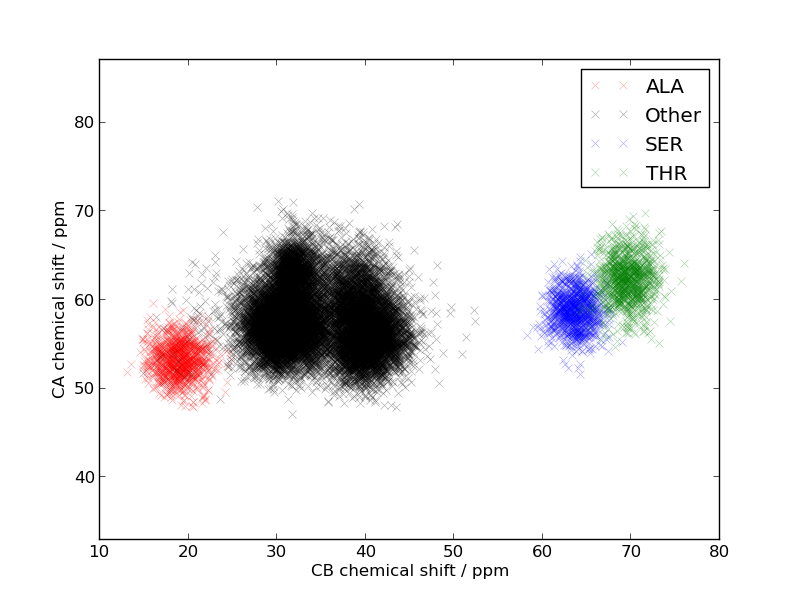}}\hspace{-20pt}
  \subfloat[]{%
    \includegraphics[width=.52\textwidth]{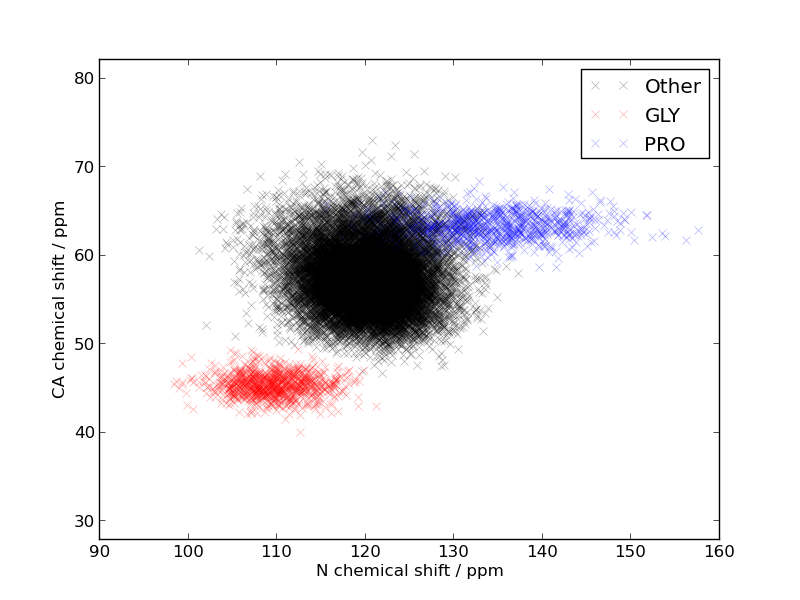}}\\
  \caption{1000 samples for each residue-type taken from normal approximations from BMRB to the distribution of chemical shifts. Residues that can't be determined near-uniquely from their chemical shifts are shown as black crosses. a) CB vs. CA chemical shifts. b) N vs CA chemical shifts.}
  \label{fig:BMRB_samples}
\end{figure}

\clearpage
\newpage

\bibliography{main}

\begin{thebibliography}{10}

\bibitem{MeilerBaker}
Jens Meiler and David Baker.
\newblock Rapid protein fold determination using unassigned nmr data.
\newblock {\em Proc. Natl. Acad. Sci. USA}, 100(26):15404–15409, 2003.

\bibitem{Vendruscolo}
Andrea Cavalli, Xavier Salvatella, Christopher~M. Dobson, and Michele
  Vendruscolo.
\newblock Protein structure determination from nmr chemical shifts.
\newblock {\em Proc. Natl. Acad. Sci. USA}, 104(23):9615--9620, 2006.

\bibitem{Bax}
Yang Shen, Oliver Lange, Frank Delaglio, Paolo Rossi, James~M. Aramini, Gaohua
  Liu, Alexander Eletsky, Yibing Wu, Kiran~K. Singarapu, Alexander Lemak,
  Alexandr Ignatchenko, Cheryl~H. Arrowsmith, Thomas Szyperski, Gaetano~T.
  Montelione, David Baker, and Ad~Bax.
\newblock Consistent blind protein structure generation from nmr chemical shift
  data.
\newblock {\em Proc. Natl. Acad. Sci. USA}, 105(12):4685--4690, 2007.

\bibitem{Garant}
Christian Bartels, Peter Güntert, Martin Billeter, and Kurt Wüthrich.
\newblock Garant-a general algorithm for resonance assignment of
  multidimensional nuclear magnetic resonance spectra.
\newblock {\em J. Comp. Chem.}, 18(1):139--149, 1998.

\bibitem{MARS}
Young-Sang Jung and Markus Zweckstetter.
\newblock Mars - robust automatic backbone assignment of proteins.
\newblock {\em J. Biomol. NMR}, 30:11--23, 2004.

\bibitem{TATAPRO}
H.S. Atreya, S.C. Sahu, K.V.R. Chary, and Girjesh Govil.
\newblock A tracked approach for automated nmr assignments in proteins
  (tatapro).
\newblock {\em J. Bio. NMR}, 17(2):125--136, 2000.

\bibitem{Autoassign}
Diane~E. Zimmerman, Casimir~A. Kulikowski, Yuanpeng Huang, Wenqing Feng,
  Mitsuru Tashiro, Sakurako Shimotakahara, Chen ya~Chien, Robert Powers, and
  Gaetano~T. Montelione.
\newblock Automated analysis of protein nmr assignments using methods from
  artificial intelligence.
\newblock {\em J. Mol. Biol.}, 269:592--610, 1997.

\bibitem{FLYA}
Elena Schmidt and Peter Güntert.
\newblock A new algorithm for reliable and general nmr resonance assignment.
\newblock {\em J. Am. Chem. Soc.}, 134:12817--12829, 2012.

\bibitem{BMRB}
Eldon~L. Ulrich, Hideo Akutsu, Jurgen~F. Doreleijers, Yoko Harano, Yannis~E.
  Ioannidis, Jundong Lin, Miron Livny, Steve Mading, Dimitri Maziuk, Zachary
  Miller, Eiichi Nakatani, Christopher~F. Schulte, David~E. Tolmie, R.~Kent
  Wenger, Hongyang Yao, and John~L. Markley.
\newblock Biomagresbank.
\newblock {\em Nucleic Acids Research}, 36:D402--D408, 2008.

\bibitem{RefDB}
Haiyan Zhang, Stephen Neal, and David~S. Wishart.
\newblock Refdb: A database of uniformly referenced protein chemical shifts.
\newblock {\em J. Biomol. NMR}, 25:173--195, 2003.

\bibitem{Sauer}
Stephan P.~A. Sauer.
\newblock {\em Molecular Electromagnetism - A Computational Chemistry
  Approach}.
\newblock Oxford University Press Inc., New York, 2011.

\bibitem{Venters}
Ronald~A. Venters, Richele Thompson, and John Cavanagh.
\newblock Current approaches for the study of large proteins by nmr.
\newblock {\em J. Mol. Struct.}, 602-603:275--292, 2002.

\bibitem{Higman:proteinNMR}
Victoria~A. Higman.
\newblock Protein nmr - a practical guide, October 2013.
\newblock http://www.protein-nmr.org.uk/.

\bibitem{ISD}
M.~Habeck, W.~Rieping, and M.~Nilges.
\newblock Weighting of experimental evidence in macromolecular structure
  determination.
\newblock {\em Proc. Natl. Acad. Sci. USA}, 103:1756--1761, 2006.

\bibitem{ISD:Science}
W.~Rieping, M.~Habeck, and M.~Nilges.
\newblock Inferential structure determination.
\newblock {\em Science}, 309(5732):303--306, 2005.

\bibitem{Olsson}
Simon Olsson, Wouter Boomsma, Jes Frellsen, Sandro Bottaro, Tim Harder, Jesper
  Ferkinghoff-Borg, and Thomas Hamelryck.
\newblock Generative probabilistic models extend the scope of inferential
  structure determination.
\newblock {\em J. Mag. Res.}, 213:182--186, 2011.

\bibitem{Jaynes1}
E.~T. Jaynes.
\newblock Information theory and statistical mechanics.
\newblock {\em Phys. Rev.}, 106(4):620--630, 1957.

\bibitem{Jaynes2}
E.~T. Jaynes.
\newblock Information theory and statistical mechanics. ii.
\newblock {\em Phys. Rev.}, 108(2):171--190, 1957.

\bibitem{TorusDBN}
Wouter Boomsma, Kanti~V. Mardia, Charles~C. Taylor, Jesper Ferkinghoff-Borg,
  and Anders Krogh.
\newblock A generative, probabilistic model of local protein structure.
\newblock {\em Proc. Natl. Acad. Sci. USA}, 105(26):8932--8937, 2008.

\bibitem{Basilisk}
Wouter~Boomsma Tim~Harder, Martin Paluszewski, Jes Frellsen, Kristoffer~E
  Johansson, and Thomas Hamelryck.
\newblock Beyond rotamers: a generative, probabilistic model of side chains in
  proteins.
\newblock {\em BMC Bioinformatics}, 11(306), 2010.

\bibitem{CLT}
Georg Pólya.
\newblock Über den zentralen grenzwertsatz der wahrscheinlichkeitsrechnung und
  das momentenproblem.
\newblock {\em Mathematische Zeitschrift}, 8(3-4):171--181, 1920.

\bibitem{SPARTA}
Yang Shen and Ad~Bax.
\newblock Protein backbone chemical shifts predicted from searching a database
  for torsion angle and sequence homology.
\newblock {\em J. Biomol. NMR}, 38(4):289--302, 2007.

\bibitem{PROSHIFT}
Jens Meiler.
\newblock Proshift: Protein chemical shift prediction using artificial neural
  networks.
\newblock {\em J. Biomol. NMR}, 26(1):25--37, 2003.

\bibitem{SHIFTX}
Stephen Neal, Alex~M. Nip, Haiyan Zhang, and David~S. Wishart.
\newblock Rapid and accurate calculation of protein 1h, 13c and 15n chemical
  shifts.
\newblock {\em J. Biomol. NMR}, 26(3):215--240, 2003.

\bibitem{Camshift}
Kai~J. Kohlhoff, Paul Robustelli, Andrea Cavalli, Xavier Salvatella, and
  Michele Vendruscolo.
\newblock Fast and accurate predictions of protein nmr chemical shifts from
  interatomic distances.
\newblock {\em J. Am. Chem. Soc}, 131(39):13894--13895, 2009.

\bibitem{Phaistos}
Wouter Boomsma, Jes Frellsen, Tim Harder, Sandro Bottaro, Kristoffer~E.
  Johansson, Pengfei Tian, Kasper Stovgaard, Christian Andreetta, Simon Olsson,
  Jan~B. Valentin, Lubomir~D. Antonov, Anders~S. Christensen, Mikael Borg,
  Jan~H. Jensen, Kresten Lindorff-Larsen, Jesper Ferkinghoff-Borg, and Thomas
  Hamelryck.
\newblock Phaistos: A framework for markov chain monte carlo simulation and
  inference of protein structure.
\newblock {\em J. Comp. Chem}, 34:1697--1705, 2013.

\bibitem{Metropolis}
Nicholas Metropolis, Arianna~W. Rosenbluth, Marshall~N. Rosenbluth, Augusta~H.
  Teller, and Edward Teller.
\newblock Equation of state calculations by fast computing machines.
\newblock {\em J. Chem. Phys.}, 21(6):1087, 1953.

\bibitem{AIC}
H~Akaike.
\newblock A new look at the statistical model identification.
\newblock {\em IEEE Transactions on Automatic Control}, 19(6):716--723, 1974.

\bibitem{AICc}
Kenneth~P. Burnham and David~R. Anderson.
\newblock Multimodel inference - understanding aic and bic in model selection.
\newblock {\em Sociological Methods and Research}, 33:261--304, 2004.

\bibitem{HNCA}
B.T.~Farmer II, R.A. Venters, L.D. Spicer, M~. G~. Wittekind, and L.~Müller.
\newblock A refocused and optimized hnca: Increased sensitivity and resolution
  in large macromolecules.
\newblock {\em J. Biomol. NMR}, 2(2):195--202, 1992.

\end{thebibliography}
\end{document}